\documentclass[runningheads, article]{llncs}

\newcommand{\ourtitle}{TweetCred: Real-Time Credibility Assessment of Content on Twitter}
\newcommand{\ourshorttitle}{TweetCred: Real-Time Credibility Assessment on Twitter}

\usepackage[show]{chato-notes}

\usepackage[square, comma, sort&compress, numbers]{natbib}
\usepackage{graphicx}
\usepackage{subfigure}
\usepackage{color}
\usepackage{paralist}
\usepackage{booktabs}
\usepackage{multirow}
\usepackage{xspace}

\usepackage[pdftex]{hyperref}
\hypersetup{
  pdftitle={\ourtitle},
  pdfauthor={},
  pdfkeywords={},
  pdfstartview={FitH},
  colorlinks, citecolor=black, filecolor=black, linkcolor=black, urlcolor=black,
  bookmarksnumbered, breaklinks=true,
}

\newcommand{\spara}[1]{\smallskip\noindent{\bf #1}}

\newcommand{\TweetCred}{\textit{TweetCred}\xspace}

\begin{document}

\title{\ourtitle}
\titlerunning{\ourshorttitle}  

\author{Aditi~Gupta\inst{1} \and Ponnurangam~Kumaraguru\inst{1} \and Carlos~Castillo\inst{2} \and Patrick~Meier\inst{2}}
\authorrunning{Aditi Gupta et al.} 
\institute{Indraprastha Institute of Information Technology, Delhi, India\\
\email{\{aditig, pk\}@iiitd.ac.in}
\and
Qatar Computing Research Institute, Doha, Qatar\\
\email{chato@acm.org, pmeier@qf.org.qa}}
\maketitle              

\begin{abstract}
During sudden onset crisis events, the presence of spam, rumors and fake content on Twitter reduces the value of information contained on its messages (or ``tweets'').
A possible solution to this problem is to use machine learning to automatically evaluate the credibility of a tweet, i.e. whether a person would deem the tweet believable or trustworthy.
This has been often framed and studied as a supervised classification problem in an off-line (post-hoc) setting.

In this paper, we present a semi-supervised ranking model for scoring tweets according to their credibility. 
This model is used in \TweetCred, a real-time system that assigns a credibility score to tweets in a user's timeline. 
\TweetCred, available as a browser plug-in, was installed and used by 1,127 Twitter users within a span of three months. During this period, the credibility score for about 5.4 million tweets was computed, allowing us to evaluate  \TweetCred in terms of response time, effectiveness and usability.
%
%
To the best of our knowledge, this is the first research work to develop a real-time system for credibility on Twitter, and to evaluate it on a user base of this size.

\end{abstract}
\section{Introduction}

Twitter is a micro-blogging web service with over 600 million users all across the globe. Twitter has gained reputation over the years as a prominent news source, often disseminating information faster than traditional news media. 
%
Researchers have shown how Twitter plays a role during crises, providing valuable information to emergency responders and the public, helping reaching out to people in need, and assisting in the coordination of relief efforts (e.g. \cite{gupta20131,Mendoza:2010:TUC:1964858.1964869,Vieweg:2010:MDT:1753326.1753486}).

On the other hand, Twitter's role in spreading rumors and fake news has been a major source of concern.
Misinformation and disinformation in social media, and particularly in Twitter, has been observed during major events that include the 2010 earthquake in Chile~\cite{Mendoza:2010:TUC:1964858.1964869}, the Hurricane Sandy in 2012~\cite{gupta2013faking} and the Boston Marathon blasts in 2013~\cite{gupta20131}.
Fake news or rumors spread quickly on Twitter and this can adversely affect thousands of people~\cite{Oh:2011:ICT:1968924.1968971}. 
Detecting credible or trustworthy information on Twitter is often a necessity, especially during crisis events.
However, deciding whether a tweet is credible or not can be difficult, particularly during a rapidly evolving situation.

Both the academic literature, which we survey on Section~\ref{sec:litreview}, and the popular press,\footnote{\url{http://www.huffingtonpost.com/dean-jayson/twitter-breaking-news_b_2592078.html}} have suggested that a possible solution is to automatically assign a score or rating to tweets, to indicate its trustworthiness.
In this paper, we introduce \TweetCred (available at \url{http://twitdigest.iiitd.edu.in/TweetCred/}), a novel, practical solution based on ranking techniques to assess {\em credibility} of content posted on Twitter in real-time. 
We understand credibility as ``the quality of being trusted and believed in,'' following the definition in the Oxford English Dictionary. A tweet is said to be credible, if a user would trust or believe that the information contained on it is true. 

\begin{figure}[t]
\centering
\fbox{\includegraphics[scale=.27]{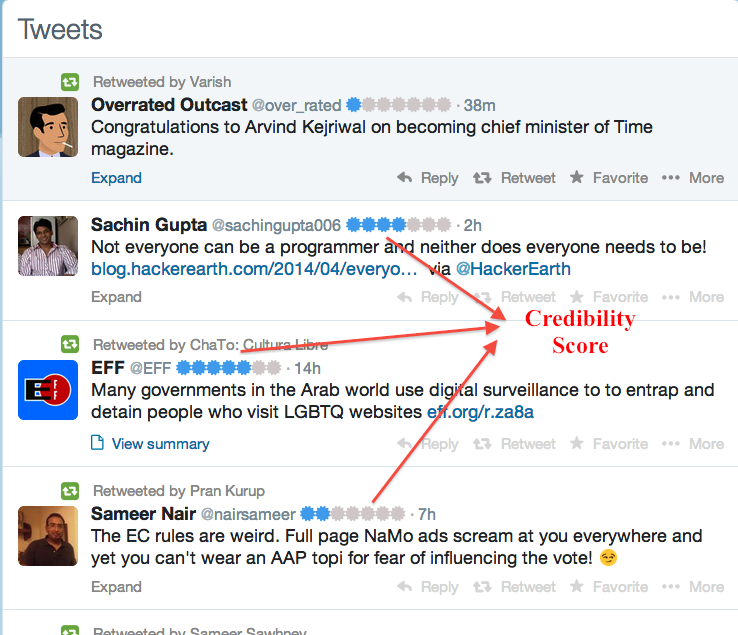}}
\caption{Screenshot of timeline of a Twitter user when \TweetCred browser extension is installed.}
\label{fig:screenshot}
\end{figure}

In contrast with previous work based on off-line classification of content in a post-hoc setting (e.g. \cite{Gupta:2012:CRT:2185354.2185356,Mendoza:2010:TUC:1964858.1964869} and many others), \TweetCred uses only the data available on each message, without assuming extensive historical or complete data for a user or an event.
Also in contrast with previous work, we evaluate \TweetCred with more than a thousand users who downloaded a browser extension that enhanced their Twitter timeline, as shown in Figure~\ref{fig:screenshot}.

The main contributions of this work are:
\begin{itemize}
\item We present a semi-supervised ranking model using SVM-rank for assessing credibility based, on training data obtained from 6 high impact crisis events of 2013. An extensive set of 45 features is used to determine the credibility score for each of the tweets. 

\item We develop and deploy a real time system, \TweetCred, in the form of a browser extension, web application, and REST API. The \TweetCred extension was installed and used by 1,127 Twitter users within a span of three months, computing the credibility score for about 5.4 million tweets.

\item We evaluate the performance of \TweetCred in terms of response time, effectiveness and usability. We observe that 80\% of the credibility scores are computed and displayed within 6 seconds, and that 63\% of users either agreed with our automatically-generated scores or disagreed by 1 or 2 points (on a scale from 1 to 7).

\end{itemize}

This paper is organized as follows:
Section~\ref{sec:litreview} briefly reviews work done around this domain.
Section~\ref{sec:methodology} describes how we collect labeled data to train our system, and
Section~\ref{sec:analysis} how we apply a learning-to-rank framework to learn to automatically rank tweets by credibility.
Section~\ref{sec:implementation} presents the implementation details and a performance evaluation, and Section~\ref{sec:user-testing} the evaluation from users and their feedback.
Finally, in the last section we discuss the results and future work.

\section{Survey}\label{sec:litreview}


%
In this section, we briefly outline some of the research work done to assess, characterize, analyze, and compute trust and credibility of content in online social media.

\spara{Credibility Assessment.}
\citet{Castillo:2011:ICT:1963405.1963500} showed that automated classification techniques can be used to detect news topics from conversational topics and assess their credibility based on various Twitter features. They achieved a precision and recall of 70-80\% using a decision-tree based algorithm.  
\citet{guptaetal} in their work on analyzing tweets posted during the terrorist bomb blasts in Mumbai (India, 2011), showed that the majority of sources of information are unknown and have low Twitter reputation (small number of followers). 
The authors in a follow up study applied machine learning algorithms (SVM-rank) and information retrieval techniques (relevance feedback) to assess credibility of content on Twitter~\cite{Gupta:2012:CRT:2185354.2185356}, finding that 
only 17\% of the total tweets posted about the event contained situational awareness information that was credible. 
Another, similar work was done by \citet{xia} on tweets generated during the England riots of 2011. They used a supervised method based on a Bayesian Network to predict the credibility of tweets in emergency situations. \citet{Donovan} focused their work on finding indicators of credibility during different situations (8 separate event tweets were considered). Their results showed that the best indicators of credibility were URLs, mentions, retweets and tweet length.

\spara{Credibility perceptions.}
\citet{Morris:2012:TBU:2145204.2145274} 
conducted a survey to understand users' perceptions regarding credibility of content on Twitter. 
They found that the prominent features based on which users judge credibility are features visible at a glance, for example, the username and picture of a user. 
Yang et al.~\cite{Yang:2013:MCP:2441776.2441841} analyzed credibility perceptions of users on two micro-blogging websites: Twitter in the USA and Weibo in China. They found that location and network overlap features had the most influence in determining the credibility perceptions of users. 

\spara{Credibility of users.}
\citet{caniniSP11} analyzed the usage of automated ranking strategies to measure credibility of sources of information on Twitter for any given topic. The authors define a credible information source as one which has trust and domain expertise associated with it. 
Ghosh et al.~\cite{Ghosh:2012:CCS:2348283.2348361} identified topic-based experts on Twitter using features obtained from user-created list, relying on the wisdom of Twitter's crowds.

\spara{System.}
\citet{ratkiewicz:truthy} introduced Truthy,\footnote{http://truthy.indiana.edu/} a system to study information diffusion on Twitter and compute a trustworthiness score for a public stream of micro-blogging updates related to an event. Their focus is to detect political smears, astroturfing, and other forms of politically-motivated disinformation campaigns.

\medskip
To the best our knowledge, the work presented in this paper is the first research work that describes the creation and deployment of a practical system for credibility on Twitter, including the evaluation of such system with real users.

\section{Training Data Collection}\label{sec:methodology}

\TweetCred is based on semi-supervised learning. As such, it requires as input a {\em training set} of tweets for which a credibility label is known.

To create this training set, we collect data from Twitter using Twitter's streaming API,\footnote{\url{https://dev.twitter.com/docs/api/streaming}} filtering it using keywords representing six prominent events in 2013: 
\begin{inparaenum}[(i)]
\item the Boston Marathon blasts in the US,
\item Typhoon Haiyan/Yolanda in the Philippines,
\item Cyclone Phailin in India,
\item the shootings in the Washington Navy Yard in the US,
\item a polar vortex cold wave in North America, and 
\item the tornado season in Oklahoma, US.
\end{inparaenum}
These events affected a large population and generated a high volume of content in Twitter. Table~\ref{tab:summary-training-data} describes the characteristics of the data collected around the events we used to build a training set.

\begin{table}[t]
\caption{Number of tweets and distinct Twitter users from which data was collected for the purposes of creating a training set. From each event, 500 tweets were labeled.}
\label{tab:summary-training-data}
\centering\small\begin{tabular}{lrr}\\
\toprule
{Event}&Tweets&Users\\\midrule
Boston Marathon Blasts &7,888,374 &3,677,531\\
Typhoon Haiyan / Yolanda & 671,918 &368,269\\
Cyclone Phailin & 76,136 &34,776\\
Washington Navy yard shootings & 484,609 &257,682\\
Polar vortex cold wave & 143,959 &116,141\\
Oklahoma Tornadoes & 809,154 &542,049\\\midrule
  Total  &10,074,150&4,996,448 \\\bottomrule
  \end{tabular}%
\end{table}

In order to create ground truth for building our model for credibility assessment, we obtained labels for around 500 tweets selected uniformly at random from each event.
The annotations were obtained through crowdsourcing provider CrowdFlower.\footnote{\url{http://www.crowdflower.com/}} We selected only annotators living in the United States. For each tweet, we collected labels from three different annotators, keeping the majority among the options chosen by them.

The annotation proceeded in two steps.
In the first step, 
we asked users if the tweet contained information about the event to which it corresponded, with the following options:
\begin{compactenum}[{---R}1.]
\item The tweet contains information about the event.
\item The tweet is related to the event, but contains no information.
\item The tweet is not related to the event.
\item None of the above (skip tweet).
\end {compactenum}
Along with the tweets for each event, we provided a brief description of the event and links from where users could read more about it. 
In this first step, 45\% of the tweets were considered informative (class R1), while 40\% were found to be related to the event for which they were extracted, but not informative (class R2), and 15\% were considered as unrelated to it (class R3).

In the second step, we selected the 45\% of tweets that were marked as informative, and annotated them with respect to the credibility of the information conveyed by it. We provided a definition of credibility (``the quality of being trusted and believed in''), and example tweets for each option in the annotation. We asked workers to score each tweet according to its credibility with the following options:
\begin{compactenum}[{---C}1.]
\item Definitely credible.
\item Seems credible.
\item Definitely incredible.
\item None of the above (skip tweet).
\end {compactenum}
Among the informative tweets, 52\% of tweets were labeled as \emph{definitively credible}, 35\% as \emph{seems credible}, and 13\% as \emph{definitively incredible}.

\section{Credibility Modeling}\label{sec:analysis}

Our aim is to develop a model for ranking tweets by credibility. We adopt a semi-supervised learning-to-rank approach. First, we perform feature extraction from the tweets. Second, we compare the speed and accuracy of different machine learning schemes, using the training labels obtained in the previous section.

\subsection{Feature Extraction}

Generating feature vectors from the tweets is a key step that impacts the accuracy of any statistical model built from this data. We use a collection of features from previous work~\cite{Castillo:2011:ICT:1963405.1963500,AnupamaAggarwal:2012kx,Gupta:2012:CRT:2185354.2185356,yardi:spam}, restricting ourselves to those that can be derived from single tweets in real-time.

A tweet as downloaded from Twitter's API contains a series of fields in addition to the text of the message.\footnote{\url{https://dev.twitter.com/docs/api/1.1/get/search/tweets}} For instance, it includes meta-data such as posting date, and information about its author at the time of posting (e.g. his/her number of followers).
For tweets containing URLs, we enriched this data with information from the Web of Trust (WOT) reputation score.\footnote{The WOT reputation system computes website reputations using ratings received from users and information from third-party sources. \url{https://www.mywot.com/}} 
The features we used can be divided into several groups, as shown in Table~\ref{tab:features}. In total, we used 45 features.

\begin{table}[t]
\caption{Features used by the credibility model.}
\label{tab:features}%
\centering\scriptsize
    \begin{tabular}{p{.2\columnwidth}p{.78\columnwidth}}
   \toprule
Feature set & Features \\
\midrule
Tweet meta-data & Number of seconds since the tweet; Source of tweet (mobile / web/ etc); Tweet contains geo-coordinates
\\
\mbox{Tweet content} \mbox{(simple)} & Number of characters; Number of words; Number of URLs; Number of hashtags; Number of unique characters; Presence of stock symbol; Presence of happy smiley; Presence of sad smiley; Tweet contains `via'; Presence of colon symbol
\\
\mbox{Tweet content} \mbox{(linguistic)} & Presence of swear words; Presence of negative emotion words; Presence of positive emotion words; Presence of pronouns; Mention of self words in tweet (I; my; mine)
\\
Tweet author & Number of followers; friends; time since the user if on Twitter; etc.\\
Tweet network & Number of retweets; Number of mentions; Tweet is a reply; Tweet is a retweet
\\
Tweet links & WOT score for the URL; Ratio of likes / dislikes for a YouTube video\\
\bottomrule
      \end{tabular}%
\end{table}

\subsection{Learning Scheme}

We tested and evaluated multiple learning-to-rank algorithms to rank tweets by credibility. We experimented with various methods that are typically used for information retrieval tasks: Coordinate Ascent~\cite{metzler2007linear}, AdaRank~\cite{Xu:2007:ABA:1277741.1277809}, RankBoost~\cite{964285} and SVM-rank~\cite{Joachims:2002:OSE:775047.775067}.
We used two popular toolkits for ranking, RankLib\footnote{\url{http://sourceforge.net/p/lemur/wiki/RankLib/}} and SVM-rank.\footnote{\url{http://www.cs.cornell.edu/people/tj/svm_light/svm_rank.html}}

Coordinate Ascent is a standard technique for multi-variate optimization, which considers one dimension at a time.
SVM-rank is a pair-wise ranking technique that uses SVM (Support Vector Machines). It changes the input data, provided as a ranked list, into a set of ordered pairs, the (binary) class label for every pair is the order in which the elements of the pair should be ranked. 
%
AdaRank trains the model by minimizing a loss function directly defined on the performance measures. It applies a boosting technique in ranking methods. 
%
RankBoost is a boosting algorithm based on the AdaRank algorithm; it also runs for many iterations or rounds and uses boosting techniques to combine weak rankings.

\spara{Evaluation metrics.}
The two most important factors for a real-time system are correctness and response time, hence, we compared the methods based on two evaluation metrics, NDCG (Normalized Discounted Cumulative Gain) and running time.
NDCG is useful to evaluate data having multiple grades, as is the case in our setting. Given a query $q$ and its rank-ordered vector V of results  $\langle v_{1},\ldots,v_{m}\rangle $, let label($v_i$) be the judgment of $v_i$. The discounted cumulative gain of V at document cut-off value $n$ is:
$$
DCG@n=\Sigma_{i=1}^{n}\frac{1}{log_{2}(1+i)}(2^{label(v_{i})}-1)~.
$$

The normalized DCG of V is the DCG of V divided by the DCG of the ``ideal" (DCG-maximizing) permutation of V (or 1 if the ideal DCG is 0). The NDCG of the test set is the mean of the NDCGs of the queries in the test set.
%

To map the training labels from Section~\ref{sec:methodology} to numeric values, we used the following transformation: 5=Informative and definitively credible (class R1.C1), 4=Informative and seems credible (R1.C2), 3=Informative and definitively incredible (R1.C3), 2=Not informative (R2), 1=Not related (R3). From the perspective of quality of content in a tweet, a tweet that is not credible, but has some information about the event, is considered better than a non-informative tweet.

\spara{Evaluation.}
We evaluated the different ranking schemes using 4-fold cross validation on the training data.
Table~\ref{tab:ranking} shows the results. We observe that AdaRank and Coordinate Ascent perform best in terms of $NDCG@n$ among all the algorithms; SVM-rank is a close second.
The gap is less as we go deeper into the result list, which is relevant given that Twitter's user interface allow users to do ``infinite scrolling'' on their timeline, looking at potentially hundreds of tweets.

The table also presents the learning (training) and ranking (testing) times for each of the methods. The ranking time of all methods was less than one second, but the learning time for SVM-rank was, as expected, much shorter than for any of the other methods. 
Given that in future versions of \TweetCred we intend to re-train the system using feedback from users, and hence need short training times, we implemented our system using SVM-rank.

\begin{table}
\caption{Evaluating ranking algorithms in terms of Normalized Discounted Cumulative Gain (NDCG) and execution times. Boldface values in each row indicate best results.}
  \label{tab:ranking}%
  \centering\small
    \begin{tabular}{lrrrr}
 \toprule
&\multicolumn{1}{c}{AdaRank}&\multicolumn{1}{c}{Coord. Ascent}&\multicolumn{1}{c}{RankBoost}&\multicolumn{1}{c}{SVM-rank}\\\midrule
NDCG@25&{\bf 0.6773}&0.5358&0.6736&0.3951\\
NDCG@50&{\bf 0.6861}&0.5194&0.6825&0.4919\\
NDCG@75&0.6949&{\bf 0.7521}&0.6890&0.6188\\
NDCG@100 &0.6669&{\bf 0.7607}&0.6826&0.7219\\\midrule
Time (training)&35-40 secs&1 min&35-40 secs&9-10 secs\\
Time (testing)&$<$1 sec&$<$1 sec&$<$1 sec&$<$1 sec\\
\bottomrule
\end{tabular}%
\end{table}

The top 10 features for the model of credibility ranking built using SVM-Rank are:
\begin{inparaenum}[(1)]
\item tweet contains \emph{via},
\item number of characters,
\item number of unique characters,
\item number of words,
\item user has location in profile,
\item number of retweets,
\item age of tweet,
\item tweet contains a URL,
\item ratio number of statuses/followers of the author, and
\item ratio friends/followers of the author.
\end{inparaenum}
We observe that majority of the top features for assessing credibility of content were tweet based features rather user attributes.

\section{Implementation and Performance Evaluation}\label{sec:implementation}

In order to encourage many users to interact with \TweetCred, we provided it in a way that was easy to use, as a browser extension. We also provided access to \TweetCred as a web-based application and as an API, but the browser extension was much more commonly used.

\subsection{Implementation}

The implementation includes a back-end and a front-end which interact over RESTful HTTP APIs.

\begin{figure}
\centerline{\includegraphics[scale=.42]{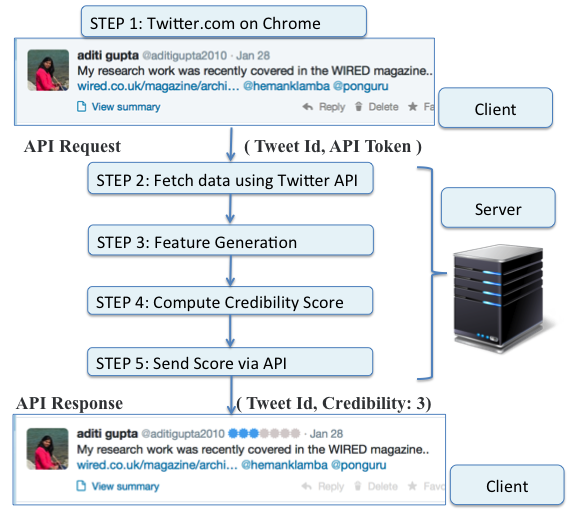}}
\caption{Data flow steps of the \TweetCred extension and API.}
\label{fig:tc}
\end{figure}

\spara{Back-end.}
Figure~\ref{fig:tc} shows the basic architecture of the system. 

The flow of information in \TweetCred is as follows: A user logs on to his/her Twitter account on \url{http://twitter.com/}, once the tweets starts loading on the webpage, the browser extension passes the IDs of tweets displayed on the page to our server on which the credibility score computation module is done.
We do not scrape the tweet or user information from the raw HTML of web page and merely pass the tweet IDs to web server. The reason is that what the server needs to compute credibility is more than what is shown through Twitter's interface.

From the server a request is made to Twitter's API to fetch the data about an individual tweet.
Once the complete data for the tweet is obtained, the feature vectors are generated for the tweet, and then the credibility score is computed using the prediction model of SVM-rank. This score is re-scaled to a value in the range from 1 to 7 using the distribution of values in our training data. Next, this score is sent back to the user's browser.
Credibility scores are cached for 15 minutes, meaning that if a user requests the score of a tweet whose score was requested less than 15 minutes ago, the previously-computed score is re-used. After this period of time, cached credibility scores are discarded and computed again if needed, to account for changes in tweet or user features such as the number of followers, retweets, favorites and replies. 

All feature extraction and credibility computation scripts were written in \emph{Python} with \emph{MySQL} as a database back-end. The RESTful APIs were implemented using \emph{PHP}. The hardware for the backend was a mid-range server (Intel Xeon E5-2640 2.50GHz, 8GB RDIMM).

\spara{Front-end.}
The Chrome browser currently enjoys the largest user base by far among various web browsers,\footnote{As of August 2014, Chrome has 59\% of market share, more than doubling the 25\% of the second place, Firefox \url{http://www.w3schools.com/browsers/browsers_stats.asp}} and hence was our target for the first version of the browser extension.
%
In order to minimize computation load on the web browser, heavy computations were offloaded to the web server, hence the browser extension had a minimalistic memory and CPU footprint. This design ensures that the system would not result in any performance bottleneck on client's web browser. 

In an initial pilot study conducted for \TweetCred with 10 computer science students that are avid Twitter users, we used the \emph{Likert Scale} of score 1--5 for showing credibility for a tweet.\footnote{http://www.clemson.edu/centers-institutes/tourism/documents/sample-scales.pdf} We collected their feedback on the credibility score displayed to them via personal interviews. The users found it difficult to differentiate between a high credibility score of 4 and a low credibility score of 2, as the difference in values seemed too small. Eight out of the ten participants felt that the scale of rankings should be slightly larger. They were more comfortable with a scale of 1--7 ranking, which we adopted.

\begin{figure}[t] 
\centering
\subfigure[]{
 \fbox{\includegraphics[width=.45\columnwidth]{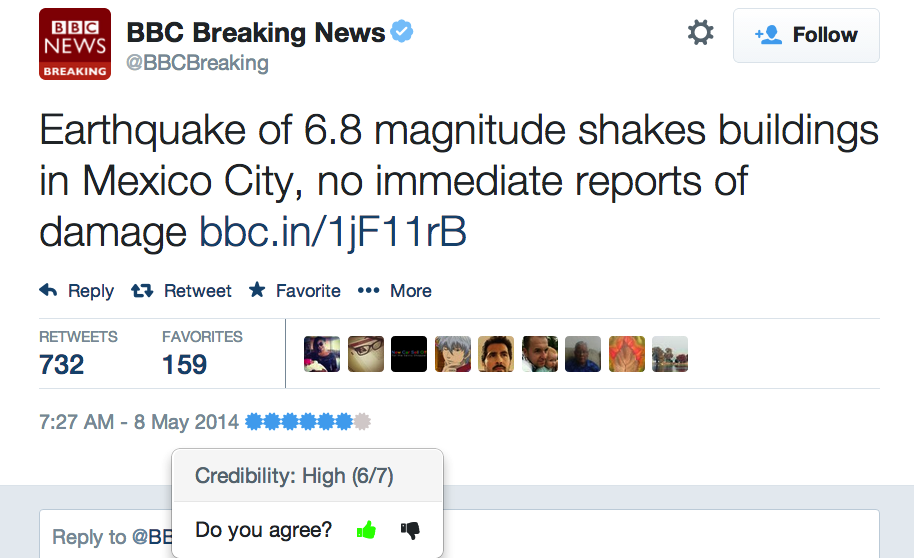}}\label{fig:agree}
}
\subfigure[]{
 \fbox{\includegraphics[width=.44\columnwidth]{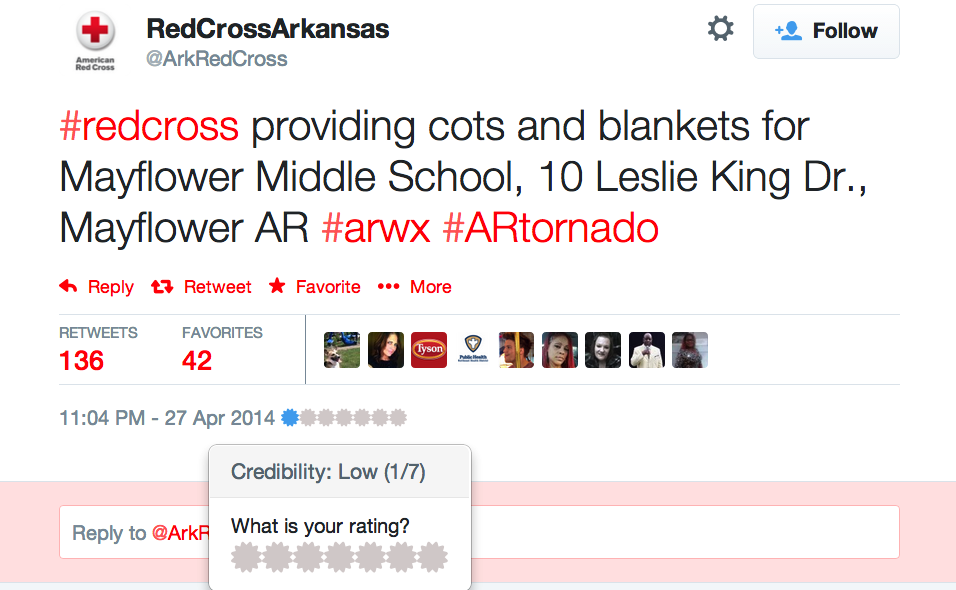}}\label{fig:mark}
}

\caption{Users can provide feedback to the system. Figure (a) shows how users can push the agree (``thumbs up'') button to agree with a rating, the case for the disagree (``thumbs down'') button is analogous. Figure (b) shows how users can provide their own credibility rating for a tweet.}
\label{fig:agree-disagree}
\end{figure}

\TweetCred displays this score next to a tweet in a user's timeline, as shown in Figure~\ref{fig:screenshot}. Additionally, the user interface includes a feedback mechanism. When end users are shown the credibility score for a tweet, they are given the option to provide feedback to the system, indicating if they agree or disagree with the credibility score for each tweet. Figures~\ref{fig:agree} shows the two options given to the user upon hovering over the displayed credibility score. In case the user disagrees with the credibility rating, s/he is asked to provide what s/he considers should be the credibility rating, as shown in Figure~\ref{fig:mark}. The feedback provided by the user is sent over a separate REST API endpoint and recorded in our database. 

\subsection{Response Time}

\begin{figure} 
\centering\includegraphics[scale=.45]{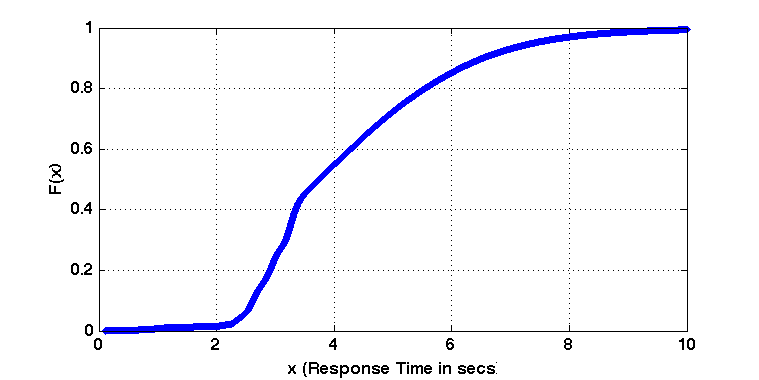}
\caption{CDF of response time of \TweetCred. For 82\% of the users, response time was less than 6 seconds and for 99\% of the users, the response time was under 10 seconds. }
\label{fig:rt}
\end{figure}

We analyzed the response time of the browser extension, measured as the elapsed time from the moment in which a request is sent to our system to the moment in which the resulting credibility score is returned by the server to the extension. Figure~\ref{fig:rt} shows the CDF of response times for 5.4 million API requests received. From the figure we can observe that for 82\% of the users the response time was less than 6 seconds, while for 99\% of the users the response time was under $10$ seconds.
The response time is dominated by the requests done to Twitter's API to obtain the details for a tweet.

\section{User Testing}\label{sec:user-testing}

We uploaded \TweetCred to the Chrome Web Store,\footnote{\url{http://bit.ly/tweetcredchrome}} and advertised its presence via social media and blogs.
We analyzed the deployment and usage activity of \TweetCred on the three-months period from April 27th, 2014 to July 31st, 2014. 
A total of $1,127$ unique Twitter users used \TweetCred. They constitute a diverse sample of Twitter users, from users having very few followers to one user having 1.4 million followers. Their usage of \TweetCred was also diverse, with two users computing the credibility scores of more than 50,000 tweets in his/her timeline, while the majority of users computed credibility scores for less than 1,000 tweets.

Table~\ref{tab:usage-stats} presents a summary of usage statistics for \TweetCred. In total 5,451,961 API requests for the credibility score of a tweet were made.

\begin{table} 
\caption{Summary statistics for the usage of \TweetCred.}
\label{tab:usage-stats}%
\centering\small
    \begin{tabular}{lr}
\toprule
Date of launch of \TweetCred& 27 Apr, 2014\\\midrule
Credibility score seen by users (total)&5,438,115\\
Credibility score seen by users (unique)&4,540,618\\
Credibility score requests for tweets (Chrome extension)& 5,429,257\\
Credibility score requests for tweets (Browser version) & 8,858 \\
Unique Twitter users& 1,127\\
\midrule
Feedback was given for tweets& 1,273\\
Unique users who gave feedback& 263\\
Unique tweets which received feedback & 1,263\\\bottomrule
  \end{tabular}%
\end{table}

\smallskip
We received feedback from users of our system in two ways. First, the users could give their feedback on each tweet for which a credibility score was computed. Secondly, we asked users to fill a usability survey on our website.

\subsection{User Feedback}

Out of the 5.4 million credibility score requests served by \TweetCred, we received feedback for 1,273 of them. When providing feedback, users had the option of either agreeing or disagreeing with our score. In case they disagreed, they were asked to mark the correct score according to them.
Table~\ref{tab:feedback} shows the break-down of the received feedback. We observed that for 40\% of tweets for which user's provided feedback agreed with the credibility score given by \TweetCred, while 60\% disagreed---this can be partially explained by self-selection bias due to cognitive dissonance: users are moved to react when they see something that does not match their expectations. 

\begin{table}[t] 
\caption{Feedback given by users of \TweetCred on specific tweets ($n=1,273$).}
\label{tab:feedback}%
\centering\small\begin{tabular}{lrr}
\toprule
& & 95\% Conf.\\
& Observed       & interval\\\midrule

Agreed with score&40.14&(36.73, 43.77)\\
Disagreed with score&59.85&(55.68, 64.26)\\\midrule
Disagreed: score should be higher &48.62&(44.86, 52.61)\\
Disagreed: score should be lower &11.23&(9.82, 13.65)\\\midrule
Disagreed by 1 point&8.71&(7.17, 10.50)\\
Disagreed by 2 points&14.29&(12.29, 16.53)\\
Disagreed by 3 points&12.80&(10.91, 14.92)\\
Disagreed by 4 points&10.91&(9.17, 12.89)\\
Disagreed by 5 points&6.52&(5.19, 8.08)\\
Disagreed by 6 points&6.59&(5.26, 8.16)\\\bottomrule
\end{tabular}
\end{table}

\spara{Credibility rating bias.}
For the approximately 60\% tweets for which users disagreed with our score, for 49\% of the tweets the users felt that credibility score should have been higher than the one given by \TweetCred, while for approximately 11\% thought it should have been lower.
This means \TweetCred tends to produce credibility scores that are lower than what users expect. This may be in part due to the mapping from training data labels to numeric values, in which tweets that were labeled as ``not relevant'' or ``not related'' to a crisis situation were assigned lower scores.
To test this hypothesis, we use keyword matches to sub-sample, from the tweets for which a credibility score was requested by users, three datasets corresponding to crisis events that occurred during the deployment of TweetCred: the crisis in Ukraine ($3,637$ tweets), the Oklahoma/Arkansas tornadoes ($1,362$ tweets), and an earthquake in Mexico ($1,476$ tweets). 

\begin{figure} 
\centering
\includegraphics[scale=.45]{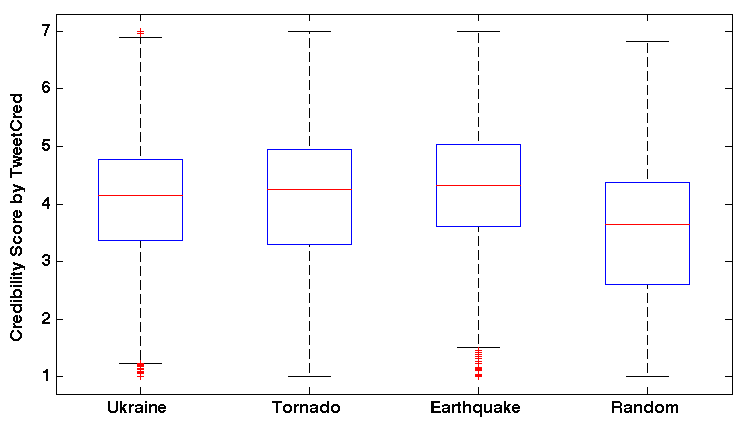}
\caption{Distribution of credibility scores. We observe that during crisis events larger percentage of tweets have higher credibility than during non-crisis.}
\label{fig:dist-credibility-scores}
\end{figure}

Figure~\ref{fig:dist-credibility-scores} compares the distribution of scores computed in real-time by TweetCred for the tweets on these three crisis events against a random sample of all tweets for which credibility scores were computed during the same time period. We observe that in all crisis events the credibility scores are higher than in the background distribution. This confirms the hypothesis that \TweetCred gives higher credibility scores to tweets that are related to a crisis over general tweets.

\subsection{Usability Survey}

To assess the overall utility and usability of the \TweetCred browser extension, we conducted an online survey among its users. An unobtrusive link to the survey appeared on the right corner of Chrome's address bar when users visited Twitter.\footnote{\url{http://twitdigest.iiitd.edu.in/TweetCred/feedback.html}} The survey link was accessible only to those users who had installed the extension, this was done to ensure that only actual users of the system gave their feedback. A total of 67 users participated.
The survey contained the standard 10 questions of the \emph{System Usability Scale} (SUS)~\cite{citeulike:4325476}. In addition to SUS questions, we also added questions about users' demographics such as gender, age, etc. We obtained an overall SUS score of 70 for \TweetCred, which is considered above average from a system's usability perspective.\footnote{\url{http://www.measuringusability.com/sus.php}} In the survey, 74\% of the users found \TweetCred easy to use (agree/strongly agree); 23\% of the users thought there were inconsistencies in the system (agree/strongly agree); and 81\% said that they may like to use \TweetCred in their daily life.

\spara{User comments.}
\TweetCred system was appreciated by majority of users for its novelty and ease of use. Users also expressed their desire to know more about the system and its backend functionality. One recurring concern of users was related to the negative bias of the credibility scores. 
Users expressed that the credibility score given by \TweetCred were low, even for tweets from close contacts in which they fully trust. For instance, one of the user of \TweetCred said: \emph{``People who I follow, who I know are credible, get a low rating on their tweets"}. Such local friendships and trust relationships are not captured by a generalized model built on the entire Twitter space. 
Other comments we received about \TweetCred in the survey and from tweets about \TweetCred were:

\begin{asparaitem}
 \item ``I plan on using this to monitor public safety situations on behalf of the City of [withheld]'s Office of Emergency Management.''
 \item ``Very clever idea but Twitter's strength is simplicity - I found this a distraction for daily use.''
 \item ``It's been good using \#TweetCred \& will stick around with it, thanks!''
 \item ``It's unclear what the 3, 4 or 5 point rating mean on opinions / jokes, versus factual statements.''
\end{asparaitem}

\section{Conclusions and Future Work}

We have described \TweetCred, a real-time web-based system to automatically evaluate the credibility of content on Twitter. 
The system provides a credibility rating from 1 (low credibility) to 7 (high credibility) for each tweet on a user's Twitter timeline. The score is computed using a semi-supervised automated ranking algorithm, trained on human labels obtained using crowdsourcing, that determines  credibility of a tweet based on more than 45 features. All features can be computed for a single tweet, and they include the tweets content, characteristics of its author, and information about external URLs.

\spara{Future work.}
Our evaluation shows that both in terms of performance, accuracy, and usability, it is possible to bring automatic credibility ratings to users on a large scale. At the same time, we can see that there are many challenges around issues including personalization and context.
With respect to personalization, users would like to incorporate into the credibility ratings the fact that their trust some of their contacts more than others. Regarding context, it is clear from the user feedback and our own observations, that there are many cases in which it may not be valid to issue a credibility rating, such as tweets that do not try to convey factual information.
In future, we would also like to study the intersection between the psychology literature about information credibility and the credibility of content in Twitter.

\TweetCred's deployment stirred a wide debate on Twitter regarding the problem and solutions for the credibility assessment problem on Twitter. The browser extension featured in many news websites and blogs including the
Washington Post,\footnote{\url{http://wapo.st/1pWE0Wd}} 
the New Yorker\footnote{\url{http://newyorker.com/online/blogs/elements/2014/05/can-tweetcred-solve-twitters-credibility-problem.html}}
and the Daily Dot\footnote{\url{http://www.dailydot.com/technology/tweetcred-chrome-extension-addon-plugin/}} among others, generating debates in these platforms. We can say that social media users expect technologies that help them evaluate the credibility of the content they read. \TweetCred is a first step towards fulfiling this expectation.

\section{Acknowledgments}
We would like to thank Nilaksh Das and Mayank Gupta in helping us with the web development of \TweetCred system. We would like to express our sincerest thanks to all members of Precog, Cybersecurity Education and Research Centre at Indraprastha Institute of Information Technology, Delhi, and Qatar Computing Research Institute for their continued feedback and support.\footnote{\url{http://precog.iiitd.edu.in/}\\ \url{http://cerc.iiitd.ac.in/}\\ \url{http://www.qcri.org.qa/}} We would like to thank the Government of India for funding this project. 

\bibliographystyle{splncsnat}
\bibliography{sigproc}




\end{document}